\documentclass[twocolumn]{aastex631}
\usepackage{amsmath}
\usepackage{longtable}
\shorttitle{Detection Limits of Thermal-Infrared Observations with Adaptive Optics}
\shortauthors{Sauter, Brandner et al.}

\begin{document}

\title{Detection Limits of Thermal-Infrared Observations with Adaptive Optics: I. Observational Data}

\author[0000-0002-8644-431X]{J. R. Sauter}
\email{sauter@mpia.de}
\affiliation{Max-Planck-Institut für Astronomie, Königstuhl 17, D-69117 Heidelberg, Germany}
\author[0000-0003-1939-6351]{W. Brandner}
\affiliation{Max-Planck-Institut für Astronomie, Königstuhl 17, D-69117 Heidelberg, Germany}
\author[0000-0002-0320-1292]{J. Heidt}
\affiliation{Landessternwarte, Zentrum für Astronomie der Universität Heidelberg, Königstuhl 12, D-69117 Heidelberg, Germany}
\author[0000-0002-3968-3780]{F. Cantalloube}
\affiliation{Université Grenoble Alpes, CNRS, IPAG, 38000 Grenoble, France}



\begin{abstract}
Ground-based thermal infrared observations face substantial challenges in correcting the predominant background emitted as thermal radiation from  the atmosphere and the telescope itself. With the upcoming 40\,m class ELTs, unprecedented sensitivities from ground will be reached, underlining the need of even more sophisticated background correction strategies. This study aims to investigate the impact of thermal backgrounds on ground-based observations and identify possible limiting factors in dedicated correction strategies. We evaluate temporal and spatial characteristics of the thermal background in direct imaging data obtained with different telescopes and observation modes. In particular, three distinct datasets, acquired using VLT/NACO and KECK/NIRC2, are analyzed. Our analysis reveals that the observations are not fully photon shot noise limited, but exhibit additional sensitivity losses caused by imperfect background compensation in the different datasets. We identify correlations between background fluctuations and the activity of the adaptive optics system. We hypothesize that the pupil modulation of the adaptive optics mirrors introduces high frequency spatial and temporal fluctuations to the background, which could ultimately constrain the detection limit if they are not compensated adequately.
\end{abstract}



\section{Introduction} \label{sec:intro}
Ground-based Thermal InfraRed (TIR) studies face a tough competition in the era of JWST. Despite the slightly superior angular resolution afforded by 8\,m to 10\,m telescopes and the availability of instrument modes not offered by JWST, such as high spectral resolution with R in the range of 10,000 to 100,000, these observations are subject to degradation by Earth's atmosphere and the emissivity of ambient temperature telescope optics and support structures.

Ground-based observations in the TIR range are confined to specific atmospheric windows ranging from 3\,µm to 25\,µm - the L-, M-, N-, and Q-bands - while being susceptible to telluric absorption and emission features. Additionally, atmospheric turbulence induces wavefront phase (and amplitude) distortions, necessitating rapid Adaptive Optics (AO) corrections for achieving diffraction-limited resolution at 8\,m-class telescopes observing in the TIR \citep{Kaeufl2018}.

Optimally designed TIR observations strive to attain Background LImited Performance (BLIP, \cite{kaufl_sky-noise_1991}), where the Signal-to-Noise Ratio (SNR) increases as SNR $\propto$ t\textsuperscript{1/2}. The time required to achieve a certain SNR is predicted to scale with telescope diameter D\textsuperscript{4} \citep{national_research_council_opticalir_1991}, implying that transitioning from an 8\,m primary mirror diameter of VLT to 39\,m of ELT should yield the same SNR in approximately 1/500 of the exposure time. Thus, the L-band detection limit achieved by instruments like NACO \citep{rousset_naos_2003,lenzen_naos-conica_2003} or ERIS \citep{davies_enhanced_2023} at the VLT in 1 hour should be reached with METIS \citep{brandl_metis_2012} at the ELT in less than 10 seconds of exposure time. However, achieving significantly fainter detection limits requires thorough calibration and removal of various atmospheric and instrumental signatures imprinted on the data. 

In this paper we aim to analyze temporal and spatial characteristics of the background signal imprinted in ground-based L'-band data obtained with VLT/NACO and KECK/NIRC2. The goal is to find and characterize limitations applied by the thermal background. In a second paper we derive a model for the thermal background, revise the implication on detection limits, and aim to devise optimized observing and data analysis strategies for future ground-based TIR studies. 

The outline of this paper is as follows:
The data and basic data reduction are described in Section \ref{sec:data}. In Section \ref{sec:detection limits} we determine the detection limits within the datasets motivating a more detailed analysis of the thermal background in Section \ref{sec:CIRB}. Here,  we investigate the background characteristics and reveal correlations with observational and instrumental parameters. We finish the paper with a general discussion and conclusion in Section \ref{sec:discussion}.

\section{Data} \label{sec:data}
We analyze two VLT/NACO datasets (hereafter referred to as the 2004 and 2011 datasets) from the ESO Science Archive Facility\footnote{\url{https://archive.eso.org/eso/eso_archive_main.html}} and an auxiliary third KECK/NIRC2 dataset from the Keck Observatory Archive\footnote{\url{https://koa.ipac.caltech.edu/cgi-bin/KOA/nph-KOAlogin}}. All three datasets comprise observations in the L'-band ($\lambda$ = 3.8\,µm, $\Delta\lambda$ = 0.7\,µm).

The VLT/NACO datasets include observations of the F8V star AF Lep, which has recently been found to have a Jovian companion \citep{de_rosa_direct_2023,franson_astrometric_2023,mesa_af_2023}. The data were recorded with a 1024 x 1024\,pixel\textsuperscript{2} Aladdin detector at an image scale of 27.1\,mas/pixel, resulting in a Field-of-View (FoV) of 28"$\times$28".

The 2004 dataset comprises a total of 48 minutes (96 $\times$ 30\,s exposures) observation time obtained over two consecutive nights (2004-12-15 and 2004-12-16). The data were acquired using random dithering and 33°-discrete Angular Differential Imaging (ADI, \citep{marois_angular_2006}) applied halfway through each individual night. The data were initially published by \cite{kasper_novel_2007}.

The 2011 dataset consists of 52 minutes (156 $\times$ 20\,s exposures) observation time obtained on 2011-10-21. The data were acquired using a four-point dithering pattern in pupil-stabilized mode, in a central 512 $\times$ 512 pixel\textsuperscript{2} sub-window, and the cube mode of VLT/NACO. The four-point dither pattern includes two pairs of observations, where two subsequent exposures are observed at the same nod position. After each pair, the nod position is moved clockwise to the next quadrant within the detector. The pupil stabilization facilitates ADI. The ADI sequence covers a total rotation angle of 70°. In the cube mode, all integrations per exposure (100 $\times$ 0.2\,s integrations) are saved. The data were initially published by \cite{rameau_survey_2013}.

The VLT/NACO datasets include AO telemetry in form of the voltage covariance matrices applied to the deformable mirror and its tip-tilt mount as well as the wavefront sensor slope covariance matrices.

The KECK/NIRC2 dataset includes observations of the A9/F0 star HIP 39017 and its companion \citep{tobin_direct-imaging_2024}. The data were recorded with a 1024 $\times$ 1024 pixel\textsuperscript{2} Aladdin detector at an image scale of 10.0\,mas/pixel, resulting in a FoV of 10" $\times$ 10". It comprises a total of 45 minutes (100 $\times$ 27 s exposures) of observation time observed on 2022-03-21. The data were obtained in pupil tracking ADI mode with a total rotation angle of 44°. It is the only utilized dataset using a coronagraph. No chopping or nodding was applied during the observations. The data were initially published by \cite{tobin_direct-imaging_2024}.

For all three datasets we use twilight flat-fields for the data reduction obtained at the same night at multiple airmasses.

The datasets are ideal for our studies as they do not include chopping, minimizing effects from non-common light paths. Utilizing datasets with different instrumental setups (i.e. discrete and continuous ADI, nodding and no nodding, or coronagraph and no coronagraph) allows us to compare and therefore distinguish the influence of the individual setups to the observed data. 

\subsection{Observing Conditions}\label{subsec:observing conditions}
The observing conditions for the VLT/NACO 2004 data were good during the first half of the first night and the entire second night, but poor during the second half of the first night. The 2004 observations were obtained at airmass ranging from 1.09 to 1.27 and experienced very low ambient wind speeds between 0.6 m/s and 3.7 m/s and ambient temperatures between 10.5°C 14.0°C.

In contrast, the conditions for the VLT/NACO 2011 data were generally moderate and highly variable. The observations occurred at smaller airmass, ranging from 1.03 to 1.04, and were accompanied by higher wind speeds ranging between 9.7 m/s and 13.2 m/s and ambient temperatures between 9.4°C and 10.0°C.

For the KECK/NIRC2 2022 dataset only the average seeing is provided, indicating only moderate observing conditions. The observations occurred at small airmass, ranging from 1.10 to 1.17 with moderate wind speeds between 1 m/s and 8 m/s and ambient temperatures between -1.0°C and -0.2°C.

A summary of the seeing conditions for the three datasets is provided in Table \ref{tab:observing conditions}.

\begin{table*}[ht]
\centering
\caption{Atmospheric seeing conditions during the VLT/NACO 2004, VLT/NACO 2011, and KECK/NIRC2 2022 datasets. The columns represent the first and second halves of the first night in 2004, the second night in 2004, the night in 2011, and 2022. The rows indicate the corresponding median values, where the errors give the peak-to-valley range. The following parameters are given (at 500 nm wavelength): seeing ($\varepsilon$), coherence length ($r_0$), coherence time ($\tau_0$), Strehl ratio ($\mathcal{S}$), and background brightness $L'_{b}$. \label{tab:observing conditions}}
\begin{tabular}{lccccc}
 Data & $\varepsilon$ ["] & $r_0$ [cm] & $\tau_0$ [ms] & $\mathcal{S}$ [\%] & \text{$L'_{b}$ [$mag/as^{2}$]}\\
 \hline
 2004.1.1 & $0.66^{+0.39}_{-0.08}$ & $17.0^{+11.4}_{-6.8}$ & $9.1^{+3.6}_{-3.1}$ & $49.0^{+13.6}_{-9.0}$ &
 $3.347^{+0.004}_{-0.003}$\\
 2004.1.2 & $1.43^{+0.62}_{-0.46}$ & $8.0^{+5.0}_{-3.2}$ & $5.7^{+4.3}_{-2.1}$ & $31.2^{+17.4}_{-30.9}$ &
 $3.343^{+0.002}_{-0.001}$\\
 2004.2 & $0.66^{+0.42}_{-0.17}$ & $15.5^{+9.8}_{-8.4}$ & $9.2^{+6.7}_{-6.4}$ & $42.4^{+8.7}_{-13.2}$ &
 $3.358^{+0.010}_{-0.016}$\\
 2011 & $1.21^{+0.63}_{-0.31}$ & $8.5^{+12.1}_{-5.7}$ & $3.0^{+4.7}_{-2.2}$ & $35.3^{+11.0}_{-33.4}$ & 
 $3.292^{+0.007}_{-0.007}$\\
 2022 & $1.08$ & $9.4$ & - & - & $1.984^{+0.008}_{-0.011}$\\
\end{tabular}
\end{table*}

\subsection{Data Reduction} \label{subsec:data reduction}
We apply a standard data reduction using bias and dark current subtraction as well as flat-field corrections to minimize pixel-to-pixel and large scale intensity variations. To isolate background effects originating from the atmosphere and telescope, we refrain from using additional reduction methods such as centroiding or background derotation, which primarily aid in stellar light subtraction and have minimal impact on subtracted backgrounds.

When subtracting two subsequent exposures to reveal temporal changes in the background, the subtracted backgrounds in the image pairs are in the following referred to as residual backgrounds. For the 2011 dataset the residual backgrounds are computed for exposures with stellar targets at the same detector position, effectively minimizing potential influences from nodding and residual stellar signals.

The stellar targets in the data are masked and then 3-$\sigma$ clipping over all non-masked pixels is used to eliminate outliers.

\section{Thermal Background Detection Limits}\label{sec:detection limits}

To assess the impact of the thermal background on ground-based observations, we calculate the 5-$\sigma$ detection limits $\mathcal{L}_{5\sigma}$ at accumulated observation time $t$ for the three datasets, using the background variance $\eta$ of the exposures as noise term in the standard SNR estimation (e.g. \cite{newberry_signal--noise_1991}):
\begin{align}
    \mathcal{L}_{5\sigma}(t,r) = \frac{25+\sqrt{625+100\cdot\eta(t)\cdot\pi r^{2}}}{2\cdot\xi(r)},
\end{align}
where $\xi$ the normalized encircled power of the underlying Point-Spread Function (PSF) at radial distance $r$.

The background variances are measured at the outer regions of the exposures, distant from the stellar locations, ensuring minimal contamination. We set the radial distance to 4\,pixel, i.e. twice the Nyquist sampling, accounting for the specific PSF shape of each dataset. It is important to note that the derived absolute limits are sensitive to the assumed PSF shape and radial detection distance, hereby providing only rough estimates in this work. The focus of our analysis lies more on the relative changes within one dataset and the conclusions that can be drawn from them. The respective utilized PSF shapes are measured using the calibration observation of each dataset and are depicted in Appendix \ref{app1}.

\begin{figure}[ht]
    \centering
    \includegraphics[width=0.45\textwidth]{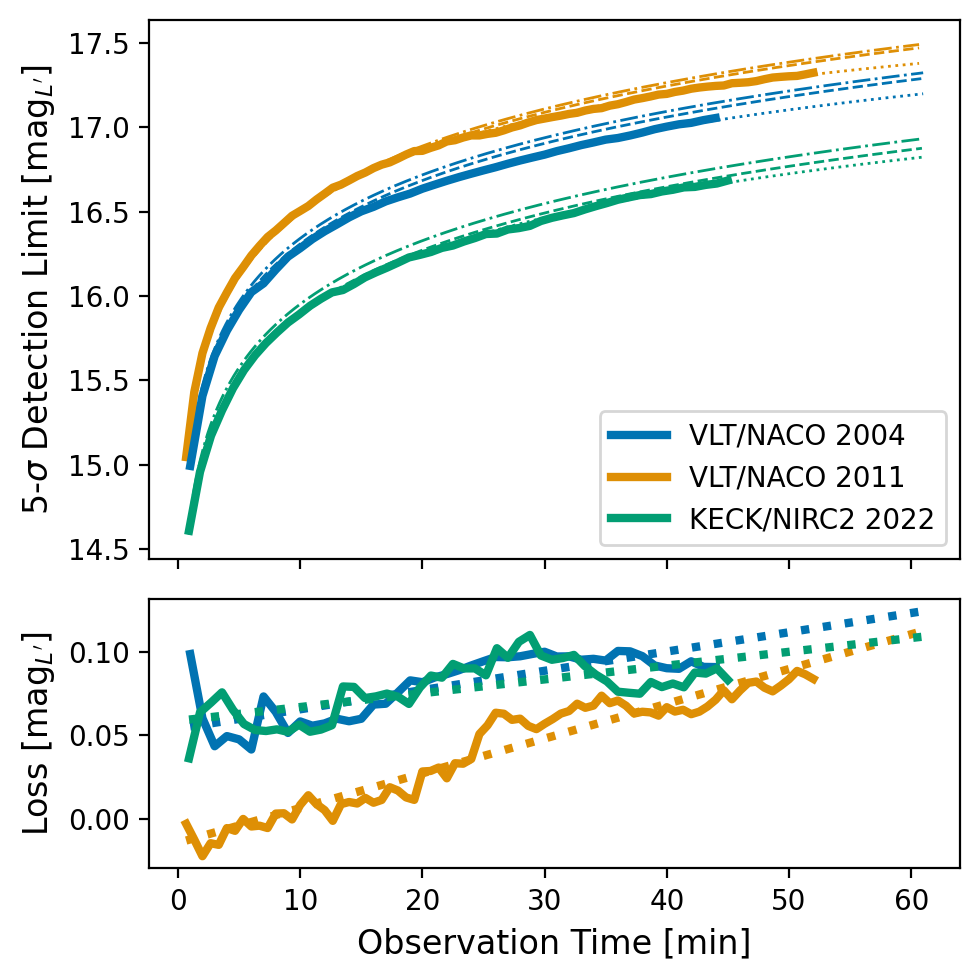}
    \caption{Top: Residual background variance 5-$\sigma$ detection limits as a function of observation time for the three different datasets. The solid lines depict the measured limits, the dashed lines the median variance limits, the dash-dotted lines the ideal median background brightness limits, and the dotted lines the median limits with applied losses from an applied linear fit. Bottom: Detection losses between the ideal median background brightness and measured detection limits (solid line). Linear fits (dotted lines) are applied to the different losses showing an overall increase with observation time.}
    \label{fig:detection limits}
\end{figure}

Figure \ref{fig:detection limits} illustrates the derived detection limits as a function of the cumulative observations time. The respective background variances are hereby either set as the measured background intensities, thus inferring ideal photon shot noise, or measured Residual Background Variance (RBV). 
The derived limits of the three datasets follow qualitatively the measured background brightness (see Table \ref{tab:observing conditions}), where higher limits are derived for the VLT/NACO datasets which comprise a lower background brightness of 3.3\,mag\textsubscript{L'}/as\textsuperscript{2} and the lower limits are derived for the KECK/NICR2 dataset which comprises a higher background brightness of 2.0\,mag\textsubscript{L'}/as\textsuperscript{2}.

The measured VLT/NACO background brightness roughly matches with the 3.0\,mag\textsubscript{L'}/as\textsuperscript{2} denoted in the VLT/NACO user manual\footnote{\url{https://www.eso.org/sci/facilities/paranal/decommissioned/naco/doc.html}}. The KECK/NIRC2 sky brightness of $\approx$2.0\,mag\textsubscript{L'}/as\textsuperscript{2} is in the range of typical background values quoted in the user manual and instrument web page\footnote{\url{https://www2.keck.hawaii.edu/inst/nirc2/ObserversManual.html} and \url{https://www2.keck.hawaii.edu/inst/nirc2/sensLong.html}}. The differences in the derived detection limits between the two VLT/NACO datasets are related to the assumed PSF shape. For the KECK/NIRC2 dataset the segmented primary mirror and pupil alignment offsets might explain the higher background brightness.

All three datasets show prominent and strongly variable excess noise with differences between the median background brightness and median variance limits of 0.03 mag\textsubscript{L'} for the 2004 dataset, 0.02 mag\textsubscript{L'} for the 2011 dataset, and 0.06\,mag\textsubscript{L'} for the 2022 dataset. Oppositely, the background brightness shows almost no variability as depicted in Table \ref{tab:observing conditions}.  A comparison between the background variance using the measured RBV and background brightness is shown in Appendix \ref{app2}.

The measured losses in all three datasets show increasing deviations from the ideal median background brightness limits with time. After approximately 50 minutes they range between 0.08\,mag\textsubscript{L'} and 0.09\,mag\textsubscript{L'}. The measured loss offsets at zero observation time result from the excess observed for the first exposures, which are relatively high for the 2004 and 2022 dataset as shown in the Appendix \ref{app2}. We apply linear fits to the measured losses to depict the linear trends in the individual datasets. Especially the 2011 dataset which depicts strongly variable seeing conditions during the observations follows its trend very precisely. Notably, the losses of the 2004 dataset tend to stay constant or even decrease in the first half of the first night and in the second night, where the seeing conditions were stable, and tend to strongly increase at the second half of the first night, where the seeing conditions were poor and very variable. The losses for the 2022 dataset show only a small slope, again qualitatively matching the average seeing conditions during the observation.

The fitted linear slopes thus represent the average seeing conditions during the observations. Although very speculative, the increasing losses would ultimately constrain the achievable detection limits when extrapolated to several hours of observation time. The applied fits would hereby imply limits of 17.8\,mag\textsubscript{L'} for the VLT/NACO datasets and 17.6\,mag\textsubscript{L'} for the KECK/NIRC2 dataset.

The measured $\sim$50\,min detection limits of 17.1\,mag\textsubscript{L'} and 17.3\,mag\textsubscript{L'} for the 2004 and 2011 VLT/NACO and 16.7\,mag\textsubscript{L'} for KECK/NIRC2 dataset surpass the corresponding planet brightness of 14.87\,mag\textsubscript{L'} for AF Lep b \citep{franson_astrometric_2023} and 16.12 mag\textsubscript{L'} for HIP 39017 b \citep{tobin_direct-imaging_2024} and would thus permit the detection in the respective datasets. Nevertheless, the derived limits clearly show the dominant role of the thermal background in achievable sensitivities. The revealed additional losses and the resulting potential saturation in sensitivity hence result in a substantial limitation for ground-based observations.

\section{Thermal Background Information Content}\label{sec:CIRB}

An ideal thermal background correction eliminates all spatial and temporal structures, leaving behind only pure, uncorrelated photon shot noise. Following \cite{larkin_reflections_2016}, intensity structures in the residual background carry valuable information or signals which should be distinguished from the pure photon shot noise. The challenge lies in interpreting this information and uncovering causal relationships. In this study, we adopt a systematic approach: First, we develop a quantitative method to measure the spatial intensity structures in the background. Then, we explore temporal correlations with ambient and instrumental parameters collected concurrently to the observations. Finally, we evaluate the correlations by comparing the different datasets to reveal possible causations.  

\subsection{Spatial Correlations}\label{subsec:spatial correlations}

\begin{figure}[ht]
    \centering
    \includegraphics[width=0.45\textwidth]{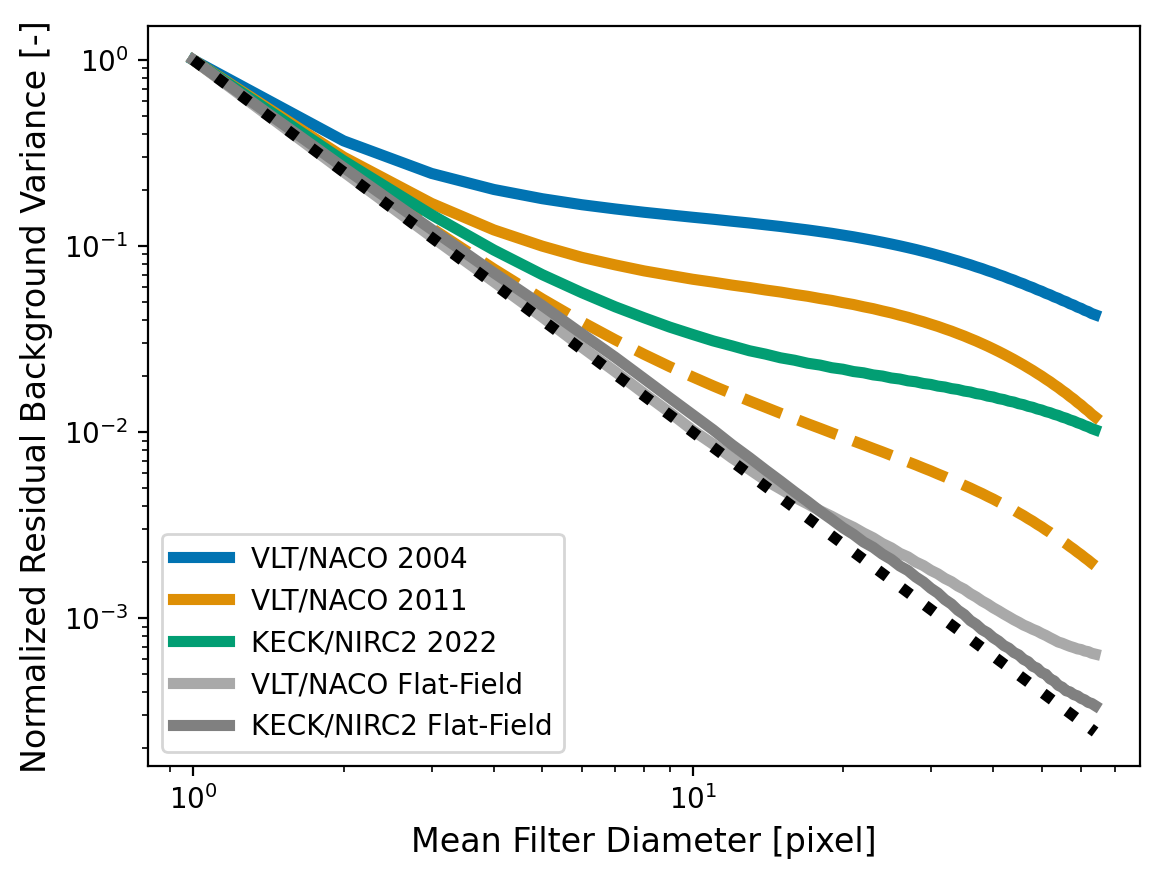}
    \caption{Normalized mean residual background variance as a function of applied mean filter diameter for the different datasets. The solid lines denote the subtracted exposure pairs, the dashed line the VLT/NACO 2011 integration pairs, and the dotted line ideal photon shot noise.}
    \label{fig:flat backgorund}
\end{figure}

The spatial intensity structure abundance of the residual backgrounds in the VLT/NACO and KECK/NIRC2 datasets, compared to those using the 2011 and KECK/NIRC2 flat-field images, are depicted in Figure \ref{fig:flat backgorund}. To characterize these structures, we measure the RBV within a 64 $\times$ 64 pixel\textsuperscript{2} box positioned far outside the stellar Airy disk at a fixed detector position. At large enough distances, the influence of diffracted stellar light - in particular by the coronagraph in the KECK/NIRC2 data - is expected to be negligible \citep{roddier_stellar_1997}. The variances are evaluated after applying a mean filter of variable diameter $d_{filter}$. Deviations from an ideal, non-correlated photon shot noise $d_{filter}^{-2}$-drop  reveal systematic correlations or structures within the background. The variances are measured for each subsequent exposure pair (and integration pair for the 2011 dataset) and are then averaged for each mean filter diameter and normalized by the variance without any filter. The individual curves thereby depict the average spatial intensity change in the timescales of one exposure (one integration) for each dataset.

From Figure \ref{fig:flat backgorund}, it is apparent that the data variances exhibit significant and qualitatively similar deviations from the flat-field variances, which almost ideally drop with increasing filter diameter. Notably, the deviations change depending on the location where the variances are measured on the detector. For the 2011 dataset, where the individual 0.2\,s integrations are available, less prominent deviations are visible, showing that the changes in the background appear already in sub-second timescales and seem to accumulate with time. From the figure it is thus evident that residual spatial intensity structures are prominent at all scales of the respective PSF shapes for all three datasets, possible influencing the achievable detection limits. The markedly different statistical properties between flat-field and science images can be solely attributed to the influence of the AO, as it is the only sub-instrument deactivated during flat-field observations that acts on theses fast time scales.

In particular, the 2011 exposures are subtracted at the same nodding position, and in the KECK/NIRC2 dataset no chopping or nodding is applied, ruling out changes in the optical path. Additionally, the 2004 dataset employs discrete ADI without angular changes between the exposures, which excludes influences of the field rotation. The twilight flat-fields were observed at identical exposure times under similar instrumental settings, atmospheric conditions, and airmass, thereby ruling out detector or atmospheric effects.

\subsection{Temporal Correlations with Ambient Parameters} \label{subsec:Correlation Ambient}
\begin{figure*}[ht]
    \centering
    \includegraphics[width=0.9\textwidth]{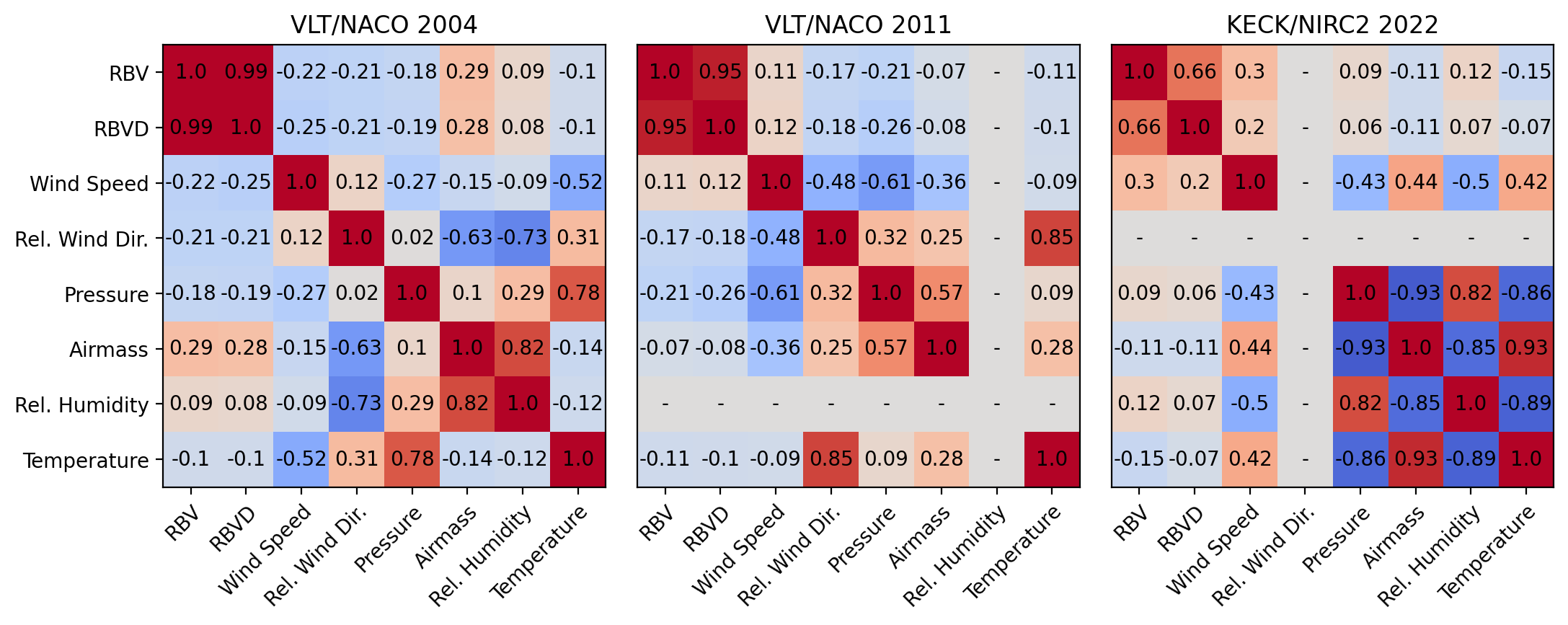}
    \caption{Normalized cross-correlation maps for the three datasets displaying the correlation between the  Residual Background Variance (RBV), the RBV Difference (RBVD), and various ambient parameters. The parameters are described in more detail in the text. The KECK/NIRC2 data lacks data on the relative wind direction, while the VLT/NACO 2011 humidity measurements did not change over the course of the observing sequence.}
    \label{fig:ambient parameter}
\end{figure*}

To analyze the temporal correlations of the background with ambient parameters, we approximate the spatial structure abundance by computing the difference between the RBV at a 10-pixel mean filter diameter and the theoretical ideal photon shot noise background variance for the different exposures of the three datasets. This parameter is denoted as RBV Difference (RBVD) and is also compared to the absolute RBV (i.e. without applying the mean filter). Figure \ref{fig:ambient parameter} displays the resulting normalized cross-correlation maps of a selection of ambient parameters: wind speed, relative wind direction, pressure, airmass, relative humidity, and temperature. The parameters are obtained from the FITS header of the different datasets (no data is available for the KECK/NIRC2 2022 wind direction, while the relative humidity remained constant at 10\% over the course of the VLT/NACO 2011 observations, thereby restricting cross-correlation). The relative wind direction is calculated as the difference between the telescope orientation and the absolute wind direction. The normalized temporal cross-correlation $\overline{\mathcal{C}}$ between two parameters $x$ and $y$ is given as:
\begin{equation}
    \overline{\mathcal{C}}(x,y) = \frac{2\cdot \mathcal{C}(\mathcal{Z}(x),\mathcal{Z}(y))}{\mathcal{C}(\mathcal{Z}(x),\mathcal{Z}(x))+\mathcal{C}(\mathcal{Z}(y),\mathcal{Z}(y))},
\end{equation}
where $\mathcal{Z}(\cdot)$ represents the standard score.

From the correlation maps, it is evident that the RBV strongly correlates with the RBVD in all three datasets. At large RBVD values, where strong spatial intensity structures appear in the background, the overall background variance is consequently increased as well. However, no consistently strong correlation between an ambient parameter and the spatial background correlation is prominent in all three datasets.

Notably, wind speed and relative wind direction exhibit the highest absolute correlations with the background over the different datasets. The anti-correlation observed in the 2004 dataset could result from the so-called low-wind effect, as reported for the VLT \citep{milli_low_2018}. In the 2022 dataset, strong correlations are present between pressure, airmass, relative humidity, and temperature. Here, the temporal airmass progression coincidentally aligns with the other parameters, all following a similar linear trend linked to the beginning of dawn. This shows that the correlations must be interpreted carefully, as causalities can be complex and not readily apparent in the data.

\subsection{Temporal Correlations with Atmospheric and AO Parameters}
\begin{figure*}[ht]
    \centering
    \includegraphics[width=0.65\textwidth]{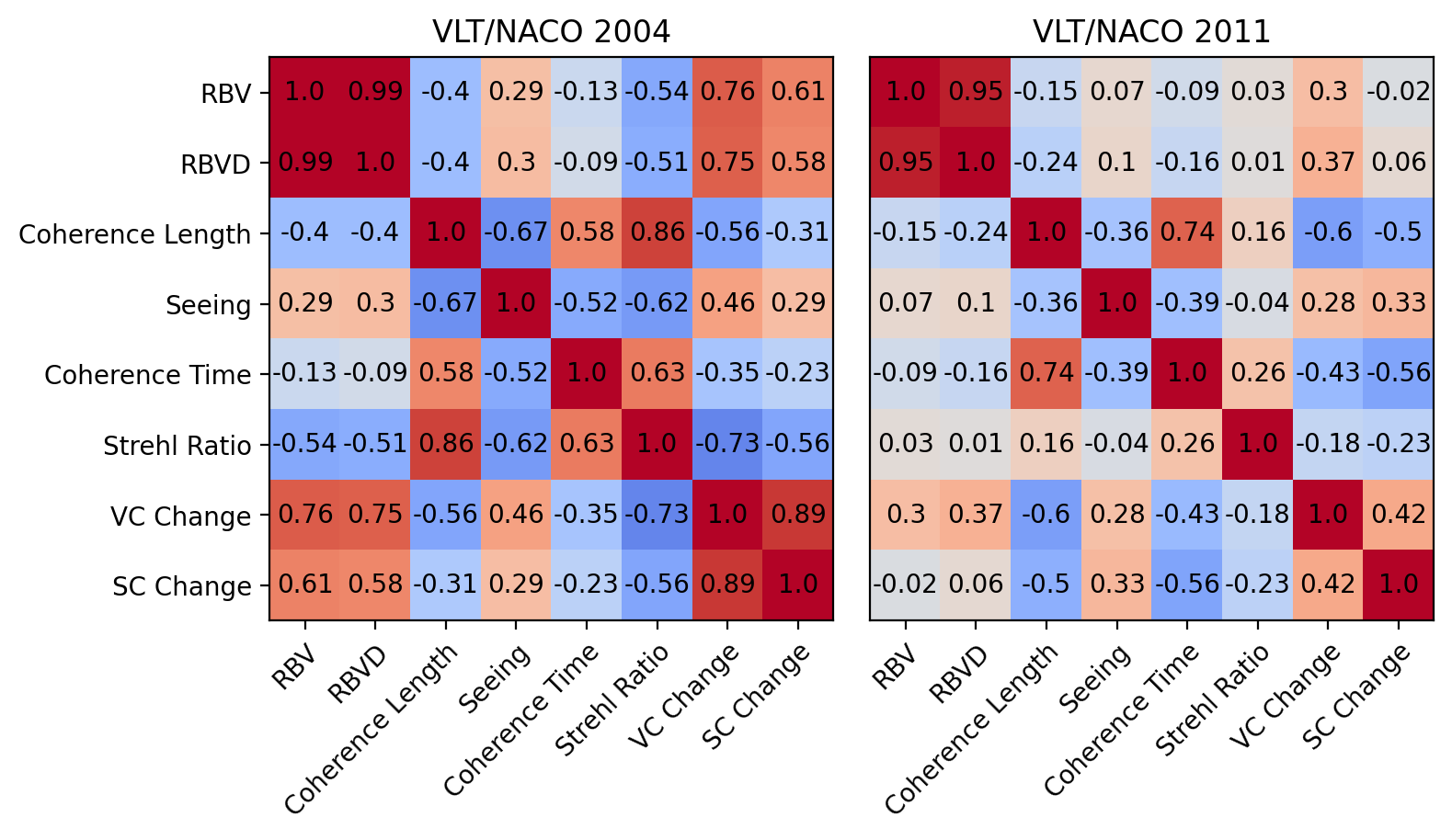}
    \caption{Normalized cross-correlation maps for the two VLT/NACO datasets displaying the correlation between the Residual Background Variance (RBV), the RBV Difference (RBVD), and various atmospheric and adaptive optics parameters. The parameters are described in more detail in the text.}
    \label{fig:atmosphere parameter}
\end{figure*}

Analogous to Section \ref{subsec:Correlation Ambient}, we investigate the temporal correlations of the background with a selection of different atmospheric and AO parameters for the two VLT/NACO datasets (no AO data is provided for the KECK/NIRC2 2022 dataset). The parameters include the coherence length, seeing, coherence time, Strehl ratio, Voltage Covariance (VC) change, and Slope Covariance (SC) change. The last two parameters denote the summed absolute differences between the corresponding covariance matrix parameters of the respective exposure pairs. The results are presented in Figure \ref{fig:atmosphere parameter}.

Compared to the ambient parameters, the atmosphere and AO parameters exhibit overall consistently strong correlations with the background structure. Among all parameters, the VC change shows the strongest correlations with the RBV and RBVD. We denote the shorter exposure times and thereby higher relative photon shot noise in the 2011 dataset to the smaller correlations between the VC change and RBV compared to the 2004 dataset. As expected, the atmospheric and AO parameters display strong correlations with each other, since worse seeing conditions are typically more variable. However, the Strehl ratio in the 2011 data deviates from this trend, possibly due to the extremely small ratios measured in this dataset. A comprehensive cross-correlation map utilizing all atmosphere, AO, and ambient parameters of the two VLT/NACO datasets can be found in Appendix \ref{app3}.

\begin{figure}[ht]
    \centering
    \includegraphics[width=0.45\textwidth]{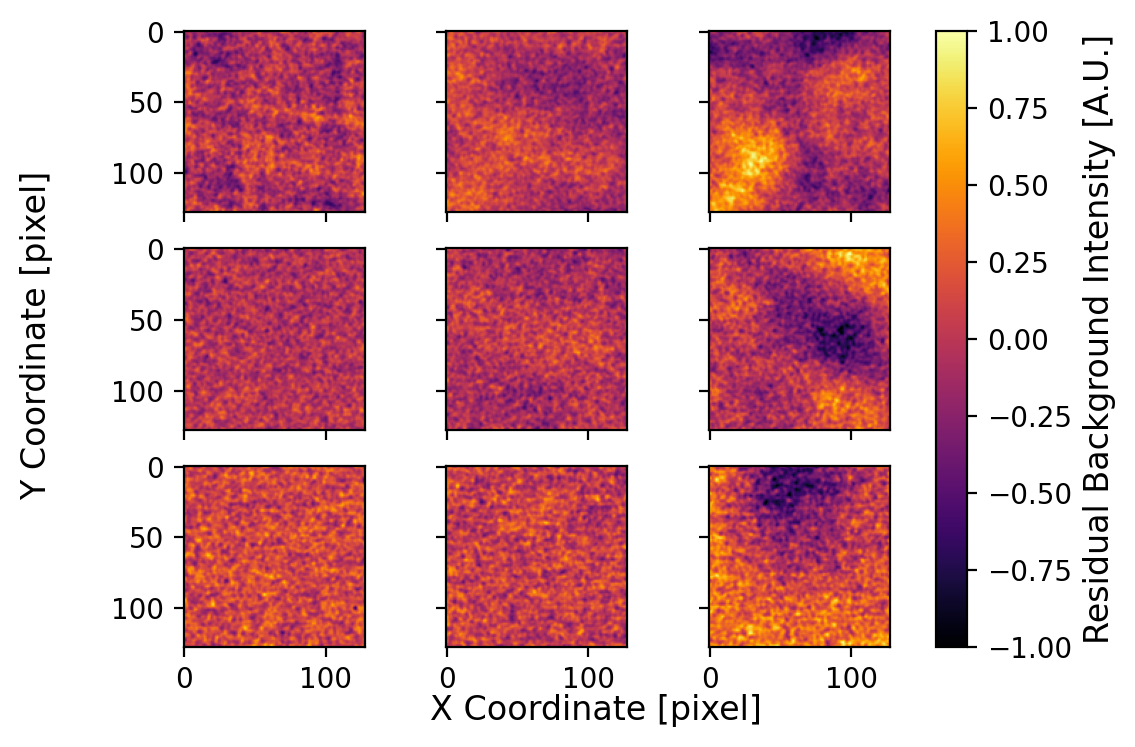}
    \caption{Example residual background cut-outs for the VLT/NACO 2004 dataset (top), VLT/NACO 2011 dataset (center), and KECK/NIRC2 dataset (bottom). The residual backgrounds for the flat-fields (left) and for low Deformable Mirror (DM) actuation (center) exhibit weak spatial intensity structures. On the other hand, the residual backgrounds for high DM actuation (right) exhibit strong spatial intensity structures. Note: the VLT/NACO 2004 flat-fields were obtained at varying rotator angles resulting in the striped features.}
    \label{fig:residual background examples}
\end{figure}

Figure \ref{fig:background voltage covariance} illustrates the relation between the RBV and the VC change in the two VLT/NACO datasets. From the figure it is evident that both parameters show a strong linear correlation. The correlation further supports our assumption in Section \ref{subsec:spatial correlations} that the AO is responsible for the spatial structures observed in the residual background.
\begin{figure}[ht]
    \centering
    \includegraphics[width=0.45\textwidth]{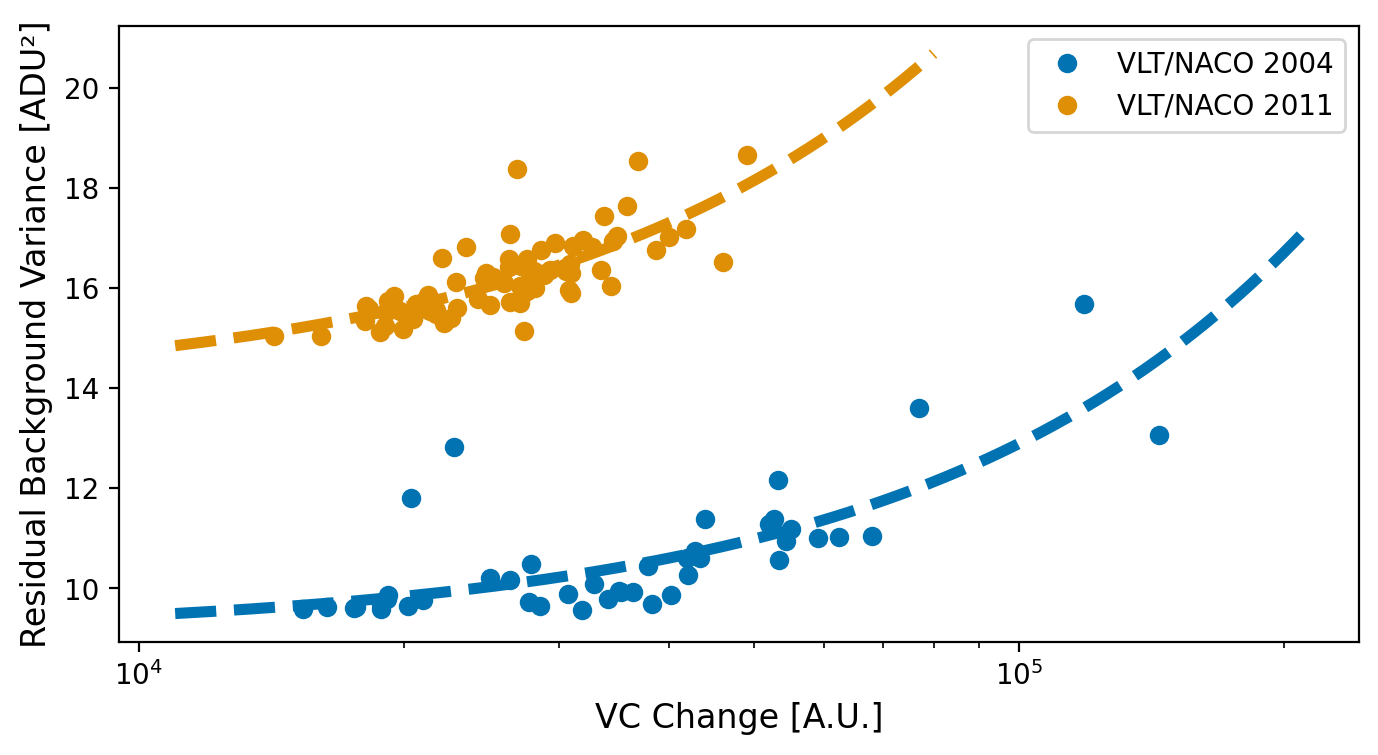}
    \includegraphics[width=0.45\textwidth]{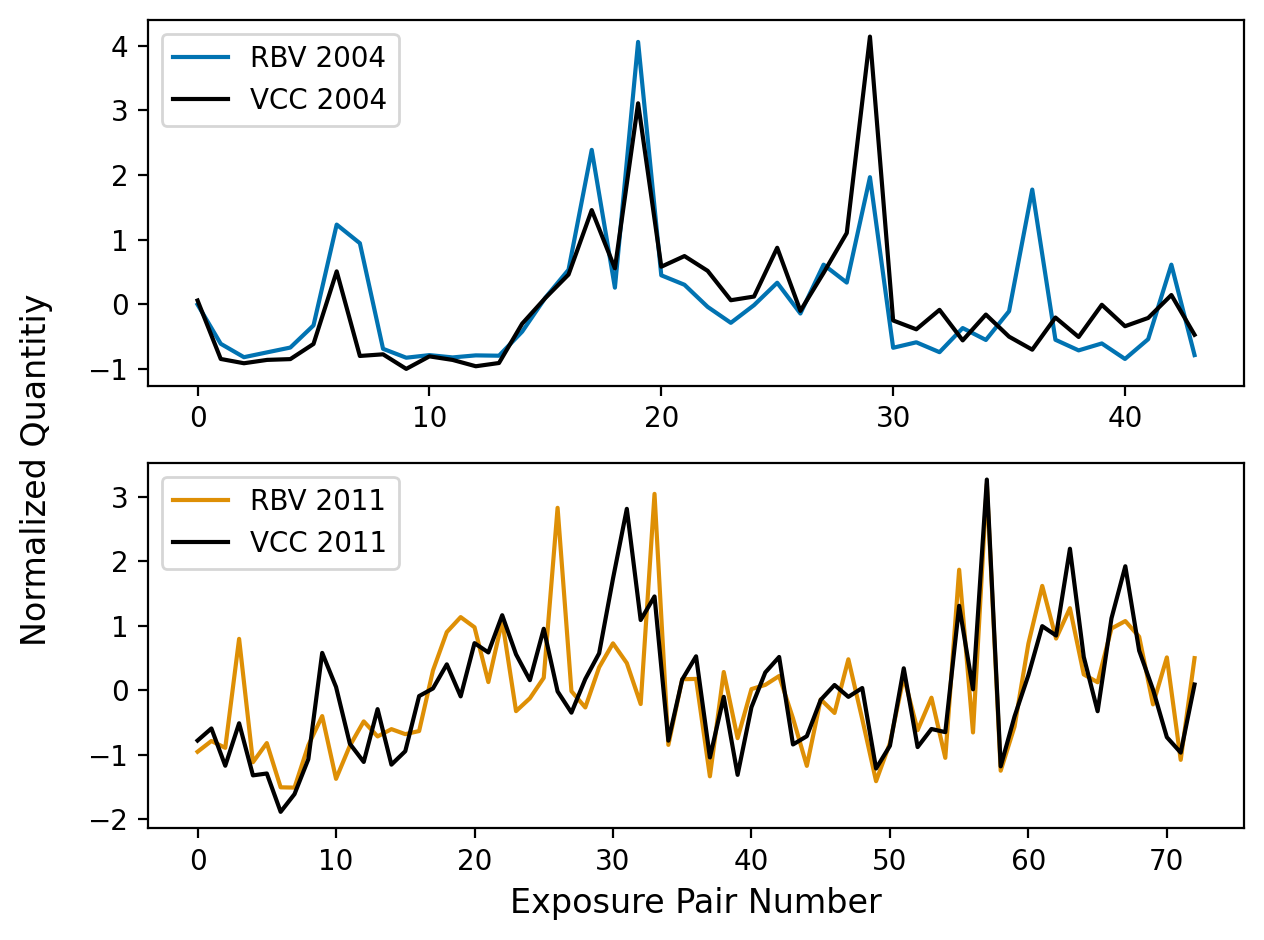}
    \caption{Top: Residual background variance as a function of voltage covariance change applied to the deformable mirror for the two VLT/NACO datasets. The dashed lines depict the linear trends within the two datasets. Bottom: The same quantities plotted individually for all exposure pairs. Note: Outliers with too high variances (2004: $>$ 22\,ADU\textsuperscript{2} and 2011: $>$ 16.5\,ADU\textsuperscript{2}) are clipped to limit the dynamic range.}
    \label{fig:background voltage covariance}
\end{figure}
In Figure \ref{fig:residual background examples} we presents example residual background cut-outs from the three datasets. For low VC changes, corresponding to low RBV, all three datasets exhibit weak spatial structures, similar to the flat-field residuals. Conversely, at high VC changes, strong spatial intensity structures become prominent. Combining the previous results, we conclude that the permanent and high-frequency changes applied by the the DM affect the background properties, resulting in spatial intensity structures after subtracting two exposures.

\section{Discussion and Conclusion} \label{sec:discussion}

The measured background detection limits in all three datasets increasingly deviate from ideal photon shot noise limits as observation time extends, resulting in a sensitivity loss of 0.08\,mag\textsubscript{L'} to 0.09\,mag\textsubscript{L'} after approximately 50 minutes of observation. The increase in loss hereby qualitatively correlates with the seeing conditions, where bad and variable seeing conditions lead to higher losses. If extrapolated to longer timescales these losses could constrain the achievable detection limits to 17.6\,mag\textsubscript{L'} to 17.8\,mag\textsubscript{L'}. These findings underscore the importance of conducting a detailed analysis to understand the origins of these losses and explore strategies to mitigate them effectively. In particular, the influence of ambient conditions, the telescope, and instruments is not well understood yet and requires further investigation.

Our examination revealed that residual backgrounds in twilight flat-field images exhibit negligible spatial intensity structures, whereas those in the science data show pronounced deviations. By comparing datasets from different telescopes and observation modes, we discovered that only the presence of AO corrections during science observations can explain this contrast. Furthermore, we identified a linear correlation between the RBV and the variability of the DM during the observations. Notably, significant background structures emerge only when the DM exhibits high variability.

These observations highlight the complex role of the AO system in background correction. We hypothesize that variations in DM configurations between exposures modulate the observed background at high frequencies, leading to systematic residuals which are responsible for the observed sensitivity losses. Further investigations are necessary to precisely delineate the causal relationship between the DM variability and background residuals and devise strategies to mitigate or counteract these effects adequately.

Expanding our analysis to include diverse telescopes, instruments, observing conditions, and atmospheric bands could further validate our findings and enhance our understanding of how factors such as telescope size, AO performance, and wavelength influence the thermal background and the applied correction outcomes. A detailed understanding is crucial to predict and maximize the potential of the next generation ELTs.

In a subsequent paper, we plan to delve deeper into these questions. Our goals include developing a model for the thermal background that attributes the causal relationship to AO and exploring the implications of background residuals on detection limits. Additionally, we aim to formulate optimized observing and data reduction strategies for future ground-based TIR studies. 

\section*{Acknowledgments}
Data obtained from the ESO Science Archive Facility under ESO programmes 074.C-0323(A) and 088.C-085(A). This research has made use of the Keck Observatory Archive (KOA), which is operated by the W. M. Keck Observatory and the NASA Exoplanet Science Institute (NExScI), under contract with the National Aeronautics and Space Administration. The KOA data were obtained under the programm U187 with T. Brandt as principal investigator.

%

\vspace{5mm}
\facilities{VLT(NACO), Keck(NIRC2)}


\software{astropy \citep{the_astropy_collaboration_astropy_2013,the_astropy_collaboration_astropy_2018,the_astropy_collaboration_astropy_2022},
Numpy \citep{harris_array_2020},
Matplotlib \citep{hunter_matplotlib_2007},
SciPy \citep{virtanen_scipy_2020}
}




\clearpage
\bibliography{References}{}
\bibliographystyle{aasjournal}

\clearpage
\appendix

\section{Appendix}\label{app1}
\begin{figure*}[ht]
    \centering
    \includegraphics[width=0.5\textwidth]{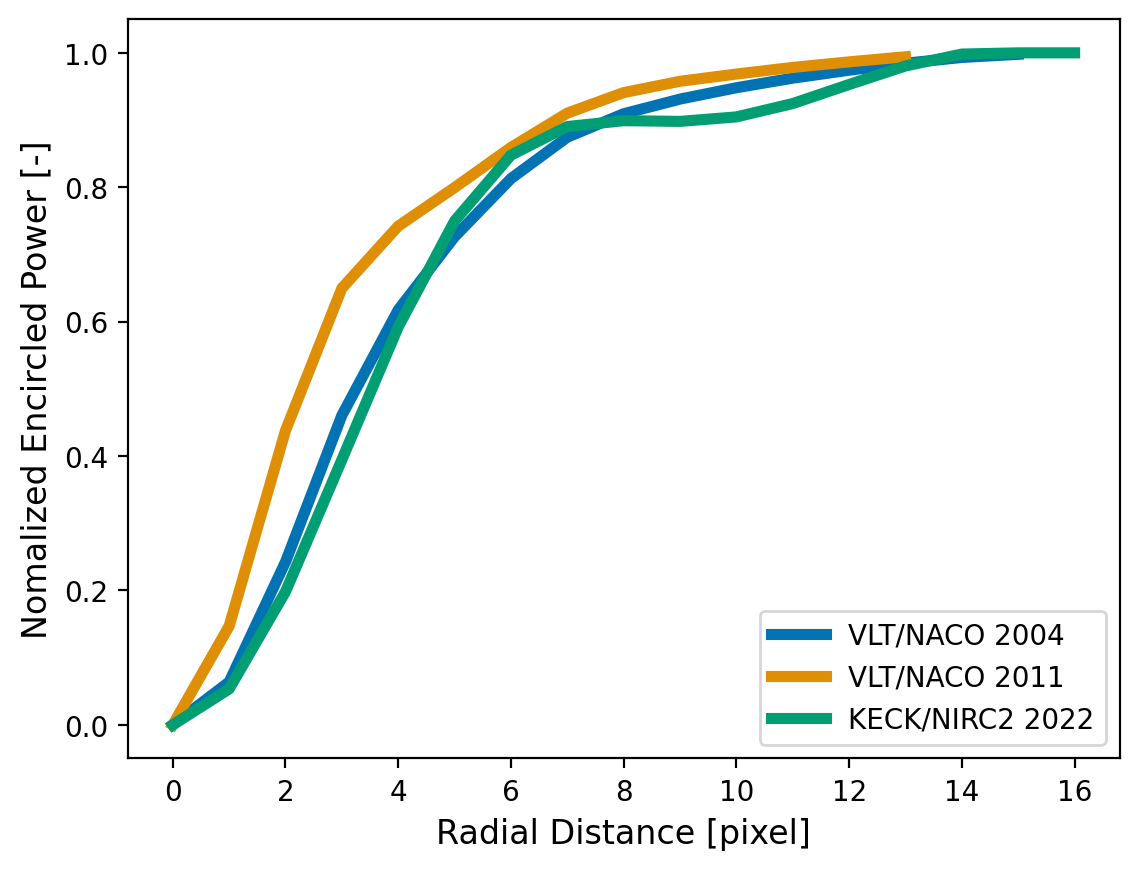}
    \caption{Normalized encircled power as a function of radial distance for the three different dataset calibration PSFs. The respective resolution is 27 mas/pixel for the VLT/NACO datasets and 10 mas/piexl for the KECK/NIRC2 dataset.}
    \label{fig:encircled power}
\end{figure*}

\section{Appendix}\label{app2}
\begin{figure*}[ht]
    \centering
    \includegraphics[width=0.5\textwidth]{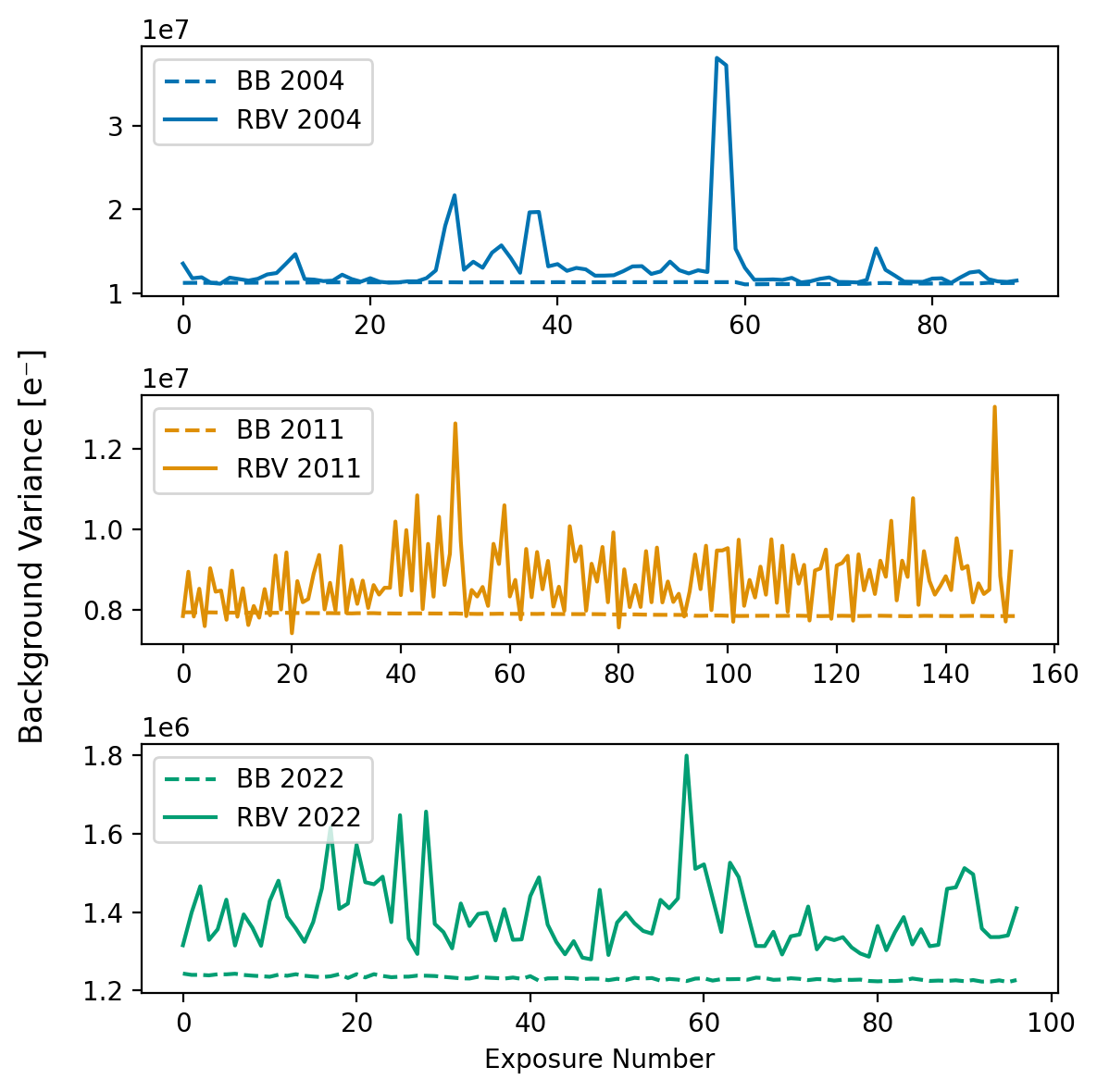}
    \caption{Derived background variance for the different exposures in the three datasets using the Background Brightness (BB) and Residual Background Variance (RBV). All three datasets exhibit variable excess noise yielding higher RBV compared to the BB.}
    \label{fig:variance exposure number}
\end{figure*}
\newpage

\section{Appendix}\label{app3}
\begin{figure*}[ht]
    \centering
    \includegraphics[width=0.55\textwidth]{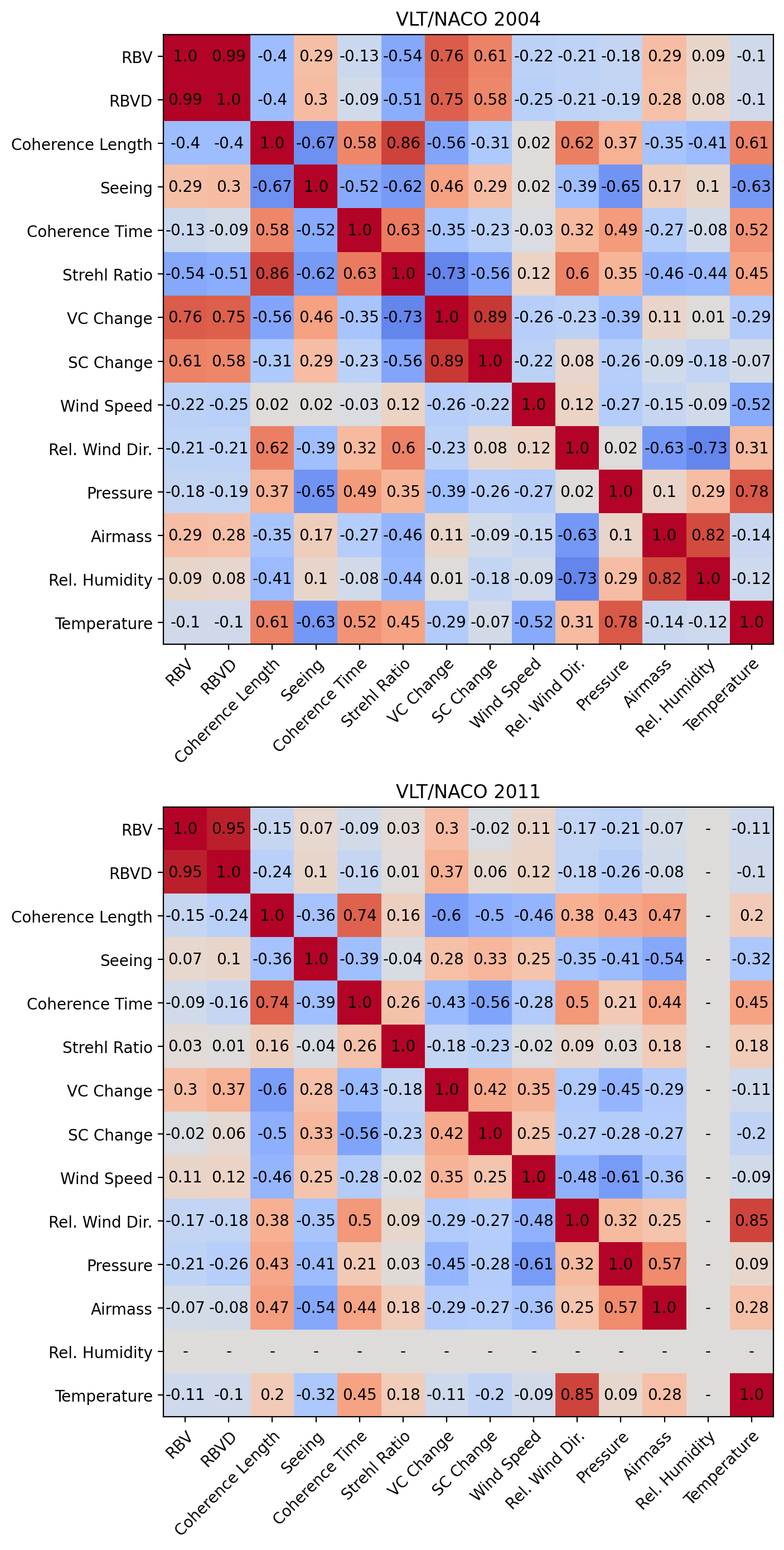}
    \caption{Normalized cross-correlation maps for the two VLT/NACO datasets displaying the correlation between the Residual Background Variance (RBV), the RBV Difference (RBVD), and different atmosphere, adaptive optics, and ambient parameters.}
    \label{fig:ambient atmosphere parameter}
\end{figure*}



\end{document}